\documentclass[12pt,draftcls,onecolumn]{IEEEtran}	

\usepackage{geometry}                		
\usepackage{bbm}
\usepackage{mathrsfs}
\usepackage{amssymb}
\usepackage{amsmath}
\usepackage{subfigure}
\usepackage{graphicx}
\usepackage{cases}
\usepackage{amsfonts}
\usepackage{amssymb}
\usepackage{booktabs}
\usepackage{multirow}
\usepackage{slashbox}
\newcommand{\bn}{\begin{eqnarray}}
\newcommand{\ed}{\end{eqnarray}}
\newcommand{\bnn}{\begin{eqnarray*}}
\newcommand{\edd}{\end{eqnarray*}}
\newcommand{\by}{\begin{array}}
\newcommand{\ey}{\end{array}}

\newcommand\old[1]{}
\newtheorem{remark}{Remark}
\newtheorem{theorem}{Theorem}
\newtheorem{lemma}{Lemma}
\newtheorem{proposition}{Proposition}
\newtheorem{definition}{Definition}

\newtheorem{assumption}{Assumption}
\newcommand{\ba}{\begin{array}}
\newcommand{\ea}{\end{array}}
\newcommand{\be}{\begin{equation}}
\newcommand{\ee}{\end{equation}}
\newcommand{\bea}{\begin{eqnarray}}
\newcommand{\eea}{\end{eqnarray}}
\usepackage[justification=centering]{caption}
\usepackage{hyperref}
\usepackage[backend=bibtex, sorting=none, bibstyle=ieee,citestyle=numeric-comp]{biblatex} 
\allowdisplaybreaks

\begin{document}
\title
{Exploration into Optimal State Estimation with Event-triggered Communication}
\author{Xiaolei Bian, Huimin Chen, X.~Rong Li, \IEEEmembership{Fellow, IEEE}

\thanks{The authors are with the Department of Electrical and Computer Engineering, University of New Orleans, New Orleans, LA 70148, U.S.A. (e-mail: xbian2@uno.edu; hchen@uno.edu; xli@uno.edu)}
}

\maketitle

\begin{abstract}
This paper deals with the problem of remote estimation of the state of a discrete-time stochastic linear system observed by a sensor with computational capacity to calculate local estimates. We design an event-triggered communication (ETC) scheme and a remote state estimator to optimally calibrate the tradeoff between system performance and limited communication resources. The novel communication scheme is the time-varying thresholding version for the cumulative innovation-driven communication scheme in \cite{bian2018remote}, and its transmission probability is given. We derive the corresponding remote minimum mean square error (MMSE) estimator and present a tight upper bound for its MSE matrices. We also show that by employing a couple of weak assumptions, the optimality problem becomes (asymptotically) exact and can be addressed in an Markov Decision Process (MDP) framework, which delivers optimal policy and cost in an algorithmic procedure. The simulation results illustrate the effectiveness of our approach.
\end{abstract}


\section{Introduction}

Wired/wireless networks in recent days are getting an ever increasing usage in control systems to enable remote estimation, monitoring and control \cite{hespanha2007survey}. These networked systems are notably strained by limited resources---energy, bandwidth and computational power \cite{park2017wireless}. With resources-awareness in mind, the community are developing ETC schemes that meet demands through intelligent allocation of resources. In ETC schemes, transmission is triggered only when certain events occur, such as the significance of data in some measure exceeding a threshold. These strategies, a.k.a threshold-based communication scheme, provide an opportunity to achieve a satisfactory estimation performance using reduced transmission since they allocate resources to important data. Such communication strategies are related to the concept of Lebesgue sampling/event-based sampling \cite{Bernhardsson1999}.

In recent years, various ETC schemes and corresponding estimators have been proposed and developed \cite{bian2018remote,Trimpe2014,yilmaz2018distributed,Han2015}.
These literature on event-based estimators either nest the design of estimators against a backdrop of ETC, or vice versa. To quantitatively account for communication resources, a cohort of researchers strive to optimally calibrate the tradeoff between estimation quality and communication expenses. The authors of \cite{imer2005optimal} considered estimating a scalar discrete-time stochastic process, with the goal of minimizing the average estimation error over a finite horizon with a limited number of transmissions. For the same objective, \cite{rabi2006multiple} studied a scalar continuous-time process. Their work was generalized to a vector linear system in \cite{li2010event} and a suboptimal solution was given. 
A related problem for an energy-harvesting sensor was considered in \cite{leong2016optimal}. Under the stochastic ETC scheme \cite{Han2015}, the optimal parameter estimation problem was solved using semidefinite programming in \cite{han2015optimal}. Similar optimization problem over an infinite horizon with a constraint on communication rate was approached to get a suboptimal solution via generalized geometric programming optimization techniques in \cite{wu2014}. The work in \cite{weimer2012distributed} introduced a distributed greedy heuristic to minimize the weighted function of the network energy consumption and the number of transmissions, subject to constraints on the estimator performance. 

Complementing constrained optimization on event-based estimation, another line of research focuses on jointly determining the optimal ETC policy and the estimator that minimize the estimation error with costly communication. The authors of \cite{lipsa2011remote} considered a first-order linear system and proved that a symmetric thresholding policy and a Kalman-like estimator are jointly optimal. In \cite{molin2017event}, an iterative algorithm was proposed that alternates optimizing state estimator and event-trigger, and both the symmetric and asymmetric thresholding policies for first-order systems were investigated. The estimation of a discrete-time Markov process with an energy-harvesting sensor was studied in \cite{nayyar2013optimal}. 
For one-dimensional, independent, and identically distributed stochastic processes, \cite{gao2018optimal} investigated the effect of communication channel noise on a similar problem and showed that the optimal communication policy is symmetric and threshold-based under some assumptions. For the stochastic ETC scheme \cite{Han2015}, the optimal threshold policy under a restricted search space was given in \cite{wu2016finite}. In \cite{leong2015optimality}, the optimal variance-based event-trigger state estimation with packet drops was considered. The aforementioned papers are concerned with finite horizon problems,  on the other hand, \cite{xu2004optimal,cogill2007constant} treats the optimization problems to minimize the weighted sum of estimation error and communication cost over an infinite horizon. The authors of \cite{xu2004optimal} adopted a Kalman-like estimator and gave the optimal ETC policy as a function of the state estimation error. A computationally tractable algorithm is given in \cite{cogill2007constant} to deliver a suboptimal policy that can guarantee a moderate upper bound of the optimal cost.  

This paper studies the optimal event-based estimation problem. We propose an ETC scheme and derive the corresponding optimal remote estimator. To quantitatively balance estimation quality and communication cost, we formulate the criterion as a weighted sum of estimation error and communication cost. To sidestep the difficulties in developing an umbrella approach in optimal estimation under a generic thresholding policy, our study bifurcates into two different courses of actions by either breaking the optimality construct while preserving the most general thresholding policy, or vice versa. Under a general thresholding policy, we derive a tight upper bound for the MSE matrices, which can be applied in a potentially suboptimal estimation problem instead. On the other hand, by respecting the optimality construct, we enforce a few assumptions which include restrictions on the class of admissible policies. We further prove that these assumptions enable an MDP architecture that produces optimal policy and cost in an algorithmic approach.

Our main contributions are as follows.

1) The ETC scheme considered in the paper is the time-varying thresholding version of the cumulative innovation-driven communication scheme in \cite{bian2018remote}. We give its transmission probability at any time. The remote MMSE estimator is derived accordingly. The information of ``no data transmission'' is used optimally to improve estimator performance. We have proved, with due mathematical rigor, that the trace of the covariance of cumulative innovation---conditioned on the information carried by the entire ``no data transmission'' duration, does not exceed the one conditioned only on the information carried by the last ``no data transmission'' timestamp. We further give a workable tight upper bound for the MSE matrices.

2) To quantify the tradeoff between estimation quality and communication expenses, we define an optimization criterion based on the expected total discounted cost---including estimation error and weighted communication cost---over the infinite horizon. We identify the decision process under our framework as an MDP one. An iterative algorithm is proposed to find the optimal policy and the optimal cost. We also extend our results to the optimal average cost problem. 

3) We prove a degree of freedom (DOF) reduction/degeneracy property that is peculiar to the optimal triggering policy. That is, the optimal thresholding decision is exclusively dependent on the tally of the elapsed time since the last transmission. No other state variables are needed in determining the threshold value at the immediate next decision point.

The paper is organized as follows. Section II formulates the remote state estimation problem and the optimization problem. Section III presents our ETC scheme and its transmission rate. Section IV derives the MMSE estimator and gives an upper bound for the trace of the MSE matrices. Section V reformats the optimization problem as an MDP problem and gives the optimal solution. Section VI presents a simulation study.
 
Notation: $\mathbb{R}$, $\mathbb{R}^{+}$, $\mathbb{Z}$, $\mathbb{N}$ denote the sets
of real numbers, positive real number, non-negative integers, and natural numbers,
respectively. A matrix $A \in {\mathbb{R}}^{n\times n}$ being positive
definite and positive semidefinite is denoted as $A > 0$ and $A \geq0$, respectively. $A^{\dagger}$ denotes the Moore–Penrose inverse of a matrix $A\in {\mathbb{R}}^{m\times n}$.
$\mathcal{N}(\mu, \Gamma)$ denotes the Gaussian distribution with mean $\mu$ and
covariance matrix $\Gamma > 0$. $\chi_{n}^{2}$ denotes the chi-square distribution with $n$ degrees of freedom.
$\{\zeta\}_{0}^{k}$ denotes the set $\{\zeta_{0}, \cdots, \zeta_{k}\}$.
$\mathbb{E}[\cdot]$ and $\mathbb{E}^{f}[\cdot]$ denote expectation, and $\textrm{Pr}(\cdot)$ denotes probability, where the underlying probability measure $\textrm{Pr}^{f}$ is parameterized by the policy $f$.
$||\cdot||$ denotes the Euclidean norm of a vector or the induced 2-norm of a matrix. $\lambda_{\text{max}}(\cdot)$ and $\textrm{tr}(\cdot)$ denote the maximum eigenvalue and the trace of a matrix, respectively. $\delta_{ij}$ denotes the Kronecker delta and $\mathbbm{1}(A)$ denotes the indicator function of a subset $A$.

\section{System model}
Consider the following discrete-time stochastic linear time-invariant system
\begin{align}
\label{eq2.1}x_{k+1}&=Ax_{k}+\omega_{k} \\
\label{eq2.2}y_{k}&=Cx_{k} +\nu_{k}, k\in \mathbb{Z}
\end{align}
where $x_{k}\in {\mathbb{R}}^{n}$ is the system state, $A\in
\mathbb{R}^{n\times n}$ ($n\in \mathbb{N}$) is the system matrix,
$y_{k}\in {\mathbb{R}}^{m}$ ($m\in \mathbb{N}$) is the measurement, $C\in \mathbb{R}^{m\times n}$ is the output matrix with rank $q$, $\omega$ is zero-mean Gaussian white process noise with covariance $Q$, and $\nu$ is zero-mean Gaussian white measurement noise with covariance $R$. The initial state is $x_{0}\sim \mathcal{N}(\bar{x}_{0},P_{0})$.
\begin{assumption}
\label{asm1} The initial state $x_{0}$, the process noise sequence $\langle w \rangle$, and the measurement
noise sequence $\langle v \rangle$ are independent.
\end{assumption}
\begin{assumption}
\label{asm2} $(A,C)$ is observable.
\end{assumption}

Suppose that the sensor generating measurements according to (\ref{eq2.2}) has  resources to find optimal local state estimates $\hat{x}_{k}^{L}$ based on
model (\ref{eq2.1})--(\ref{eq2.2}). It is well known that for a linear
Gaussian system (\ref{eq2.1})--(\ref{eq2.2}), the Kalman filter is optimal in the MMSE sense. Therefore, the optimal local state estimator is as follows
\begin{align}  
\label{eq2.4.1}\hat{x}^{L}_{k|k-1}&=\mathbb{E}[x_{k}|\{y\}_{0}^{k-1}] =A\hat{x}^{L}_{k-1}\\
 \label{eq2.4.2}P^{L}_{k|k-1}&=\mathbb{E}[(x_{k}-\hat{x}^{L}_{k|k-1})(x_{k}-\hat{x}^{L}_{k|k-1})'] =AP^{L}_{k-1}A'+Q\\
 \label{eq2.4.3}\hat{x}^{L}_{k}&=\mathbb{E}[x_{k}|\{y\}_{0}^{k}] =\hat{x}^{L}_{k|k-1}+K^{L}_{k}(y_{k}-C\hat{x}^{L}_{k|k-1})\\
 \label{eq2.4.4}P^{L}_{k}&=\mathbb{E}[(x_{k}-\hat{x}^{L}_{k})(x_{k}-\hat{x}^{L}_{k})']=P^{L}_{k|k-1}-K^{L}_{k}CP^{L}_{k|k-1}\\
 \label{eq2.4.5}K^{L}_{k}& =P^{L}_{k|k-1}C'(CP^{L}_{k|k-1}C'+R)^{-1}.
\end{align}

The local estimates $\hat{x}^{L}_{k}$ are coded into data packets and
transmitted to the estimator at the remote center via an ideal communication
channel. We aim to design a communication policy $\pi^\gamma=\{\gamma_{1}, \dots, \gamma_{k}, \dots\}$ to balance communication cost and estimation quality that respects causality. Here a binary decision variable $\gamma_{k}=1$ is used to
indicate that the sensor transmits
data at time $k$ and $\gamma_{k}=0$ otherwise. Several communication schemes have been devised for this purpose. In this paper, attention will be restricted to a type of ETC scheme---the time-varying thresholding version of the cumulative innovation-driven communication strategy, i.e., the communication policy $\pi^\gamma$ is a sequence of causal mappings of sensor measurements $\{y\}^{k}_{0}$ and transmission history $\{\gamma\}^{k-1}_{0}$.

At time $k$, the information about $x_k$ available at the remote center is denoted by $\mathcal{I}_{k}$, where $\mathcal{I}_{k}=\{\gamma_{k}\hat{x}^{L}_{k}\}\cup \{\gamma_{k}\} $.
Given $\{\mathcal{I}\}^{k}_{0}$, the MMSE estimation at the remote center is
\begin{align}  \label{eq2.5.1}
	\hat{x}_{k}&=\mathbb{E}[x_{k}|\{\mathcal{I}\}^{k}_{0}] \\\label{eq2.5.2}
	P_{k}&=\mathbb{E}[(x_{k}-\hat{x}_{k})(x_{k}-\hat{x}_{k})^{\prime }|\{%
	\mathcal{I}\}^{k}_{0}].
\end{align}

To quantitatively make a the trade-off between communication consumption and estimation quality, the cost for each time $k$ is formulated as $\mathbb{E}[||x_{k}-\hat{x}_{k}||^2+\kappa\gamma_{k}|\{\mathcal{I}\}^{k}_{0}]$, where the constant $\kappa$ is a weighted cost paid per unit of time when data is transmitted. In order to take into account the long-term effect, we consider the expected total discounted cost over the infinite horizon which is defined as follows.

Problem 1
\begin{eqnarray} \label{eq2.6.1}
&&\underset{\pi^\gamma}{\min} \underset{T \to \infty}{\lim} \mathbb{E}^{\pi^\gamma}\bigg[\sum^{T}_{k=1}\alpha^{k-1}(\mathbb{E}[||x_{k}-\hat{x}_{k}||^2+\kappa\gamma_{k}|\{\mathcal{I}\}^{k}_{0}])\bigg]
\end{eqnarray}
where $0\leq \alpha< 1$ is the discount factor. The expectations are taken with
respect to the noise processes $\omega_{k}$'s and $\nu_{k}$'s. The motivation for studying discounted problems comes mainly from a plentiful of applications, where costs in the future bear less importance than the current one. Obviously, Problem 1 can be rewritten as
\begin{eqnarray} \label{eq2.6.2}
&&\underset{\pi^\gamma}{\min} \underset{T \to \infty}{\lim} \mathbb{E}^{\pi^\gamma}\bigg[\sum^{T}_{k=1}\alpha^{k-1}(\text{tr}(P_{k})+\kappa\gamma_{k})\bigg].
\end{eqnarray}

In the reminder of the paper, we aim to answer the following questions.

1) How to design an ETC scheme and what is its transmission probability?

2) What is the MMSE estimator under our ETC scheme? How to obtain an upper bound of the MSE matrices?

3) How to solve the above optimization problem?

\section{The ETC scheme}
In this section, we propose an ETC scheme based
on the cumulative estimate innovation and time-varying thresholds. We give the transmission probability of our ETC scheme at any time.

Define the (instant) estimate innovation $z_{k}$ at time $k$ as
\begin{align}  \label{eq3.1}
z_{k}&=\mathbb{E}[x_{k}|\{y\}_{0}^{k}] -\mathbb{E}[x_{k}|\{y\}_{0}^{k-1}]=K^{L}_{k}(y_{k}-C\hat{x}^{L}_{k|k-1}).
\end{align}

\begin{lemma}
\label{lem1} The estimate innovation sequence $\{z\}_{0}^{k}$ has the following
properties:\newline
1) $z_{k} \sim \mathcal{N}(\mathbf{0}, P^{L}_{k|k-1}-P^{L}_{k})$. \newline
2) $\{z\}_{0}^{k}$ is an independent sequence.
\end{lemma}

\begin{lemma}
\label{lem2} $P^{L}_{k|k-1}-P^{L}_{k}\geq 0$ and $\mathrm{rank}(P^{L}_{k|k-1}-P^{L}_{k})=\mathrm{rank}(C)=q$.
\end{lemma}

Since $P^{L}_{k|k-1}-P^{L}_{k}\geq 0$, there exists a unitary matrix $U_{k} \in
\mathbb{R}^{n \times n}$ such that
\begin{eqnarray}
U_{k}^{\prime}(P^{L}_{k|k-1}-P^{L}_{k})U_{k}=\text{diag}(\Lambda_{k}, \mathbf{0}_{q\times (n-q)})
\end{eqnarray}
where $\Lambda_{k}=\text{diag}(\lambda_{1},\lambda_{2},\cdots, \lambda_{q})$ and $\lambda_{1},\lambda_{2},\cdots,
\lambda_{q} $ are the positive eigenvalues of $P^{L}_{k|k-1}-P^{L}_{k}$. Let $F_{k}=[(\Lambda_{k})^{-\frac{1}{2}}, \mathbf{0}_{q\times (n-q)}]U_{k}^{\prime }$. Then define the normalized (instant) estimate innovation vector
as
\begin{eqnarray}  \label{eq3.2}
\epsilon_{k}=F_{k}z_{k}.
\end{eqnarray}

Obviously, $\epsilon_{k}\sim \mathcal{N}(\mathbf{0}, I_{q})$. Without loss of generality, we will use $n$ instead of $q$ for the remainder of this paper.

For each time $k$, define the latest transmission before time $k$
\begin{eqnarray}  \label{eq3.3}
&&t_{k}=\text{max}\{t \in \mathbb{Z}: t < k, \gamma_{t}=1\}.
\end{eqnarray}

Let $\tau_{k}$ and $\tau_{k}^{+}$ denote the elapsed time of the latest transmission of before time $k$ and up to and including time $k$, respectively, i.e.,
\begin{align}  \label{eq3.4.1}
\tau_{k}&=k-t_{k},\ \ \ \ \ \ \ \ \ \ \ \ \ \ \ \ \ \ \ \ \ \ \ \ \ \ \ \ \ \ \ \ \tau_{k}\in \{1,2, \dots, k\}\\
\label{eq3.4.2}\tau_{k}^{+}&=k-\text{max}\{t \in \mathbb{Z}: t \leq k, \gamma_{t}=1\},\ \ \ \ \tau_{k}^{+}\in \{0,1, \dots, k\}.
\end{align}

When the time index is clear from the context,
we will write $t_{k}$ as $t$, $\tau_{k}$ as $\tau$ and $\tau_{k}^{+}$ as $\tau^{+}$ for simplicity. Then, it follows from (\ref{eq2.4.1}), (\ref{eq2.4.3}), (\ref{eq3.1}) that $\hat{x}_{k}^{L}$
can be represented as
\begin{align}  \label{eq3.5}
	\hat{x}_{k}^{L}&=A^{\tau}\hat{x}_{t}^{L}+A^{\tau-1}z_{t+1}+%
	\dots+z_{k}  \notag \\
	&=A^{\tau}\hat{x}^{L}_{t}+A^{\tau-1}F^{\dagger}_{t+1}\epsilon_{t+1}+\dots+F^{\dagger}_{k}\epsilon_{k}  \notag \\
	&=A^{\tau}\hat{x}^{L}_{t}+A^{F}\epsilon_{t+1,k}
\end{align}
where $F^{\dagger}_k=U_{k}[(\Lambda_{k})^{\frac{1}{2}%
}, \mathbf{0}_{q\times (n-q)}]^{\prime }$, $A^{F}=[A^{\tau-1}F^{\dagger}_{t+1},\dots, AF^{\dagger}_{k-1}, F^{\dagger}_{k}]$ and
$\epsilon_{t+1,k}=[\epsilon_{t+1}^{\prime },
\epsilon_{t+2}^{\prime }, \dots, \epsilon_{k}^{\prime }]^{\prime }$.
Obviously, $A^{F}\epsilon_{t+1,k}$ contains all innovations in the local estimate at the
current time $k$ since the latest transmission.

We propose a novel ETC scheme $\pi^{\gamma}$ based on the cumulative innovation $\epsilon_{t+1,k}$ and the time-varying threshold $\delta_{k}$. The corresponding decision variable $\gamma _{k}$ is defined as 
\begin{eqnarray}\label{eq3.6}
&&\gamma _{k}=\left\{
\begin{array}{l}
1\ \ \ \text{if}\ ||\epsilon _{t+1,k}||^{2}>\delta_{k} \\
0\ \ \ \text{if}\ ||\epsilon _{t+1,k}||^{2}\leq \delta_{k}  %
\end{array}%
\right.
\end{eqnarray}%
where $||\epsilon _{t+1,k}||^{2}$ is a cumulative innovation statistic and $\delta_{k}$ is a time-varying threshold, which can be pre-specified or designed based on a specified mathematical model of optimization. If $\gamma _{k}=1$, then the sensor sends $\hat{x}_{k}^{L}$. Information ``no data transmission''\ at time
$k$, corresponding to $||\epsilon _{t
+1,k}||^{2}\leq \delta_{k}$, can be and should be used to improve estimation
performance. Note that $||\epsilon_{t+1,k}||^2=||\epsilon_{t+1}||^{2}+||%
\epsilon_{t+2}||^{2}+\dots+||\epsilon_{k}||^2$. 

Denote the transmission probability $\text{Pr}(\gamma_{k}=1)$ as $\alpha_{k}$, $k\in \mathbb{Z}$.
\begin{lemma}
\label{lem3} Under our communication scheme, the transmission probability is given by 
\begin{align}\label{eq3.7} \nonumber
\alpha_{k}&=\text{Pr}(\chi^{2}_{n}> \delta_{k})\alpha_{k-1}+\sum_{j=2}^{k}\Bigg (\int\limits_{D_{k-j+1}}f(u_{j-1};n)\mathrm{d}u_{j-1}\cdots \int\limits_{D_{k-2}}\text{Pr}(\chi^{2}_{n}\leq \delta_{k-1}-\sum_{l=2}^{j-1}u_{l})f(u_{2};n)\mathrm{d}u_{2}\Bigg )\\
&-\int\limits_{D_{k-j+1}}f(u_{j-1};n)\mathrm{d}u_{j-1}\cdots \int\limits_{D_{k-1}}\text{Pr}(\chi^{2}_{n}\leq \delta_{k}-\sum_{l=1}^{j-1}u_{l})f(u_{1};n)\mathrm{d}u_{1} \Bigg)\alpha_{k-j}
\end{align}
where $D_{k-l}=[0,\delta_{k-l}-\sum_{h=l+1}^{j-1}u_{h}]$, $f(u_{l};n)$ is the pdf of $\chi^{2}_{n}$ ($l\in \{1,2, \cdots, j-1\}$).
\end{lemma}

\textbf{Proof:} 
\begin{align}\label{eq3.8.1}  \nonumber
\alpha_{k}&=\sum_{j=1}^{k}\text{Pr}(\gamma_{k}=1, \gamma_{k-1}=0, \cdots, \gamma_{k-j+1}=0, \gamma_{k-j}=1)\\\nonumber
&=\text{Pr}(\gamma_{k}=1 | \gamma_{k-1}=1)\alpha_{k-1}+\sum_{j=2}^{k}\text{Pr}(\gamma_{k}=1 | \gamma_{k-1}=0, \cdots, \gamma_{k-j+1}=0, \gamma_{k-j}=1)\\
&\times \text{Pr}(\gamma_{k-1}=0, \cdots, \gamma_{k-j+1}=0 | \gamma_{k-j}=1)\alpha_{k-j}
\end{align}
where
\begin{align} \label{eq3.8.2} 
&\text{Pr}(\gamma_{k}=1 | \gamma_{k-1}=1)=\text{Pr}(\chi^{2}_{n}> \delta_{k})\\\nonumber
\label{eq3.8.3} 
&\text{Pr}(\gamma_{k}=1 | \gamma_{k-1}=0, \cdots, \gamma_{k-j+1}=0, \gamma_{k-j}=1)\text{Pr}(\gamma_{k-1}=0, \cdots, \gamma_{k-j+1}=0 | \gamma_{k-j}=1)\\\nonumber
&=\Bigg (1-\frac{\text{Pr}(\gamma_{k}=0, \cdots, \gamma_{k-j+1}=0 | \gamma_{k-j}=1)}{\text{Pr}(\gamma_{k-1}=0, \cdots, \gamma_{k-j+1}=0 | \gamma_{k-j}=1)} \Bigg )\text{Pr}(\gamma_{k-1}=0, \cdots, \gamma_{k-j+1}=0 | \gamma_{k-j}=1)\\\nonumber
&=\text{Pr}(\gamma_{k-1}=0, \cdots, \gamma_{k-j+1}=0 | \gamma_{k-j}=1)-\text{Pr}(\gamma_{k}=0, \cdots, \gamma_{k-j+1}=0 | \gamma_{k-j}=1)\\\nonumber
&=\text{Pr}(||\epsilon_{l}||^{2}\leq (\delta_{l}-\sum_{h=k-j+1}^{l-1}||\epsilon_{h}||^{2}), l\in\{k-j+1,\cdots,k-1\})\\
\nonumber
&-\text{Pr}(||\epsilon_{l}||^{2}\leq (\delta_{l}-\sum_{h=k-j+1}^{l-1}||\epsilon_{h}||^{2}), l\in\{k-j+1,\cdots,k\})\\\nonumber
&=\int\limits_{D_{k-j+1}}f(u_{j-1};n)\mathrm{d}u_{j-1}\cdots \int\limits_{D_{k-2}}\text{Pr}(\chi^{2}_{n}\leq \delta_{k-1}-\sum_{l=2}^{j-1}u_{l})f(u_{2};n)\mathrm{d}u_{2}\\
&-\int\limits_{D_{k-j+1}}f(u_{j-1};n)\mathrm{d}u_{j-1}\cdots \int\limits_{D_{k-1}}\text{Pr}(\chi^{2}_{n}\leq \delta_{k}-\sum_{l=1}^{j-1}u_{l})f(u_{1};n)\mathrm{d}u_{1}.
\end{align}
By combining $(\ref{eq3.8.1})$-$(\ref{eq3.8.3})$, we obtain the transmission probability $\alpha_{k}$, $k\in \mathbb{Z}$. \ \ \ \ \ \ \ \ \ \ \ \ \  \ \ \ \ \ \ \ \ \ \ \ \ \ \ \ \ \ \ \ \ $\Box$

It can be shown that $\alpha_{k}$ is convergent as $k \to \infty$.

\section{The remote MMSE estimator}
This section shows how to obtain the remote MMSE estimator
$(\ref{eq2.5.1})$-$(\ref{eq2.5.2})$ using the sensor data under our ETC scheme.

Let the region $\Omega_{k}=\{\epsilon_{t+1,k} \in \mathbb{R}^{\tau n}:
||\epsilon_{t+1,k}||^{2} \leq \delta_{k}\}$ carry information
about the sensor data with $\gamma_{k}=0$. Note that $\gamma_{k}=0$ implies $\gamma_{i}=0\ (i\in \{t+1, \cdots, k-1, k\})$. Thus, if $\gamma_{k}=0$, then $\{\mathcal{I}\}_{0}^{k}=\{\{\mathcal{I}\}_{0}^{t}, \Omega_{t+1}, \dots, \Omega_{k-1}, \Omega_{k}\}$.

\begin{lemma}
\label{lem4}
\begin{align}
\label{eq4.3.1}&\mathbb{E}[\epsilon_{t+1,k}| \{\mathcal{I}\}_{0}^{t}, \Omega_{t+1}, \dots, \Omega_{k-1}, \Omega_{k}]=0\\\nonumber
\label{eq4.3.2}&\mathbb{E}[\epsilon_{t+1,k}(\epsilon_{t+1,k})^{'}| \{\mathcal{I}\}_{0}^{t}, \Omega_{t+1}, \dots, \Omega_{k-1}, \Omega_{k}]=\eta(\delta_{t+1}^{k})\\
&=\text{diag}(\eta_{t+1}(\delta_{t+1}^{k}), \eta_{t+2}(\delta_{t+1}^{k}), \cdots, \eta_{k}(\delta_{t+1}^{k}))
\end{align}
where $\delta_{t+1}^{k}=(\delta_{t+1}, \cdots, \delta_{k})$,\\ $\eta_{i}(\delta_{t+1}^{k})=\frac{1}{n}\times \frac{\int\limits_{D_{t+1}}f(u_{t+1};n)\mathrm{d}u_{t+1}\cdots \int\limits_{D_{i}}u_{i} f(u_{i};n)\mathrm{d}u_{i} \cdots \int\limits_{D_{k}}f(u_{k};n)\mathrm{d}u_{k}}{\int\limits_{D_{t+1}}f(u_{t+1};n)\mathrm{d}u_{t+1}\cdots \int\limits_{D_{k-1}}\text{Pr}(\chi^{2}_{n}\leq \delta_{k}-\sum_{h=t+1}^{k-1}u_{h})f(u_{k-1};n)\mathrm{d}u_{k-1}}I_{n}$, 
$D_{l}=[0,\delta_{l}-\sum_{h=t+1}^{l-1}u_{h}]$, $f(u_{l};n)$ is the pdf of $\chi^{2}_{n}$ ($i, l\in \{t+1, \cdots, k\}$). 
\end{lemma}
\textbf{Proof:} Unconditionally, $\epsilon_{k} \sim \mathcal{N}(\mathbf{0}, I_{n})$ according to Lemma \ref{lem1} and (\ref{eq3.2}). Also, $\epsilon_{t+1},
\epsilon_{t+2}, \dots, \epsilon_{k}$ are independent. Thus, unconditionally, $%
\epsilon_{t+1,k}\sim \mathcal{N}(\mathbf{0},I_{(k-t)n})$.

It is clear that $\epsilon_{t+1,k}$ and $\{\mathcal{I}\}_{0}^{t}$
are independent. Since the pdf of $\epsilon_{t+1,k}$ and the integration region, i.e., the common part of regions $\Omega_{t+1}, \cdots, \Omega_{k-1}, \Omega_{k}$ are all symmetrical about the origin, $\mathbb{E}[\epsilon_{t+1,k}| \{\mathcal{I}\}_{0}^{t}, \Omega_{t+1}, \cdots, \Omega_{k-1}, \Omega_{k}]=\mathbb{E}[\epsilon_{t+1,k}| \Omega_{t+1}, \cdots, \Omega_{k-1}, \Omega_{k}]=0.$

\begin{align}
\label{eq4.3.3}\nonumber
&\mathbb{E}[\epsilon_{t+1,k}(\epsilon_{t+1,k})^{'}| \{\mathcal{I}\}_{0}^{t}, \Omega_{t+1}, \cdots, \Omega_{k-1}, \Omega_{k}]\\
&=\left [
\begin {array} {cccc}
\epsilon_{t+1}(\epsilon_{t+1})^{'}&\dots&\epsilon_{t+1}(\epsilon_{k})^{'}\\
\vdots&\ddots&\vdots\\
\epsilon_{k}(\epsilon_{t+1})^{'}&\dots&\epsilon_{k}(\epsilon_{k})^{'}\end {array} \middle |\Omega_{t+1}, \cdots, \Omega_{k-1}, \Omega_{k}\right].
\end{align}

For $\forall i, j\in \{t+1, \cdots, k-1, k\}$, if $i \neq j$,  then $\mathbb{E}[\epsilon_{i}(\epsilon_{j})^{'}|\Omega_{t+1}, \cdots, \Omega_{k-1}, \Omega_{k}]=0$ since 1) $\epsilon_{i}$ and $\epsilon_{j}$ are both odd functions; 2) the pdf of $\epsilon_{i}$, $\epsilon_{j}$ and the integration region are symmetric around the origin.

If $i = j$,
\begin{align}\label{eq4.3.4}
&\mathbb{E}[\epsilon_{i}(\epsilon_{i})^{'}| \Omega_{t+1}, \cdots, \Omega_{k-1}, \Omega_{k}]\\\nonumber
&=\frac{1}{n}\mathbb{E}[(\epsilon_{i})^{'}\epsilon_{i}| \Omega_{t+1}, \cdots, \Omega_{k-1}, \Omega_{k}]I_{n}\\\nonumber
&=\frac{1}{n}\mathbb{E}\Bigg [||\epsilon_{i}||^{2}\Bigg |\sum_{h=t+1}^{l}||\epsilon_{h}||^{2}\leq \delta_{l}, l\in\{t+1, \cdots, k-1, k\}\Bigg ]I_{n}\\\nonumber
&=\frac{1}{n}\times \frac{\idotsint \limits_{D_{t+1}\times \cdots D_{k}} u_{i} f_{||\epsilon_{t+1}||^{2}, \cdots, ||\epsilon_{k}||^{2}}(u_{t+1}, \cdots, u_{k})\mathrm{d}u_{k}\cdots\mathrm{d}u_{t+1}}{\idotsint \limits_{D_{t+1}\times \cdots D_{k}} f_{||\epsilon_{t+1}||^{2}, \cdots, ||\epsilon_{k}||^{2}}(u_{t+1}, \cdots, u_{k})\mathrm{d}u_{k} \cdots\mathrm{d}u_{t+1}}I_{n}\\\nonumber
&=\frac{1}{n}\times \frac{\int\limits_{D_{t+1}}f(u_{t+1};n)\mathrm{d}u_{t+1}\cdots \int\limits_{D_{i}}u_{i} f(u_{i};n)\mathrm{d}u_{i} \cdots \int\limits_{D_{k}}f(u_{k};n)\mathrm{d}u_{k}}{\int\limits_{D_{t+1}}f(u_{t+1};n)\mathrm{d}u_{t+1}\cdots \int\limits_{D_{k-1}}\text{Pr}(\chi^{2}_{n}\leq \delta_{k}-\sum_{h=t+1}^{k-1}u_{h})f(u_{k-1};n)\mathrm{d}u_{k-1}}I_{n}.
\end{align}

This completes the proof.\ \ \ \ \ \ \ \ \ \ \ \ \ \ \ \ \  \ \ \ \ \ \ \ \ \ \ \ \ \ \ \ \ \ \ \ \ \ \ \ \ \ \ \ \ \  \ \ \ \ \ \ \ \ \ \ \ \ \ \ \  \ \ \ \ \ \ \ \ \ \ \ \ \ \ \ \  \ \ \ \ \ \ \ \ \ \ \ \ \ \ \ \ \ \ \ \ $\Box$
\begin{theorem}
\label{th1}With our communication scheme, under Assumption \ref{asm1}, the MMSE estimator of $x_{k}$ given $\{\mathcal{I}\}^{k}_{0}$ is
\begin{align} 
\label{eq4.4.1}\hat{x}_{k}&=(1-\gamma_{k})A\hat{x}_{k-1}+\gamma_{k}\hat{x}^{L}_{k},\\
\label{eq4.4.2}P_{k}&=P_{k}^{L}+(1-\gamma_{k})\sum_{i=0}^{\tau-1}\eta_{k-i}(\delta_{t+1}^{k})A^{i}(P^{L}_{k-i|k-i-1}-P^{L}_{k-i})(A^{i})^{
\prime }.
\end{align}
\end{theorem}
\textbf{Proof:} The proof is easy to obtain based on Theorem 1 in \cite{bian2018remote} and Lemma \ref{lem4}. \ \ \ \ \ \ \ \ \ \ \ \ \ \ \ \ \ \ $\Box$

The computation of $P_{k}$ involves numerical integration, which is often intractable in general. We try to derive an upper bound for the trace of $P_{k}$. First of all, the following definitions and lemma are given.

Let random vectors $\Xi_{1},  \Xi_{2}, \cdots, \Xi_{N}$ be i.i.d.,  and $\Xi_{j}\sim \mathcal{N}(\mathbf{0}, I_{n})$ ($j \in\{1,2, \cdots, N\}$, $N, n\in \mathbb{N}$). Denote the region $d_{j}=\{[\xi_{1}^{'},  \xi_{2}^{'}, \cdots, \xi_{N}^{'}]'\in \mathbb{R}^{Nn}:
\sum^{j}_{i=1}||\xi_{i}||^{2} \leq \delta_{j}\}$ ($ \delta_{j}\geq 0$) and  $D=d_1\cap d_2 \cdots \cap d_N$.

\begin{lemma}\label{lem5}
The boundary of $D$, denoted by $\partial D$, has the following properties:\\
1) $\partial D$ is almost everywhere smooth.\\
2) For $\forall \textbf{v} \in {\mathbb{R}}^{Nn}$, where $|\textbf{v}|=1$, $\exists a \in \mathbb{R}$, s.t. $a\textbf{v} \in \partial D$ and 
\begin{eqnarray}\label{eq4.5}
&&b\textbf{v}=\left\{
\begin{array}{l}
\in D	\ \ \ \ \ \ \ \ \ \ \ \ \ \ \ \ \ \ \ \ \ \text{if}\ b\in [0, a] \\
\notin D  \ \ \ \ \ \ \ \ \ \ \ \ \ \ \ \ \ \ \ \ \ \ \text{otherwise.}
\end{array}
\right.
\end{eqnarray}
\end{lemma}
\textbf{Proof:} 
1) Since $\partial d_{j}=\{\xi_{i}\in {\mathbb{R}}^{n}|\sum^{j}_{i=1}||\xi_{i}||^{2}=\delta_{j}\}$ are smooth and $D=d_1\cap d_2 \cdots \cap d_N$, $\partial  D$ is smooth almost everywhere.\\
2) Since $d_j$ are all closed convex sets, any intersection of any family of closed sets is closed and the intersection of any collection of convex sets is convex, we have that $D$ is a closed convex set.
Since $D$ is closed, then $\exists a \in \mathbb{R}$, s.t. $a\textbf{v} \in \partial D$, where $\textbf{v}$ is an arbitrary unit vector in ${\mathbb{R}}^{Nn}$.
Further, the convexity of $D$ delivers Eq. (\ref{eq4.5}).  \ \ \ \ \ \ \ \ \ \ \ \ \ \ \ \ \ \ \ \ \ \ \ \ \ $\Box$

Lemma \ref{lem5} shows that the region of no-data-transmission has a well-behaved boundary in the infinitesimal cone at an arbitrary angle. This property helps delivering the following result.

\begin{theorem}
\label{th2}
The following inequality holds:
\begin{eqnarray}\label{eq4.6}
&&E\Bigg[\sum^{N}_{i=1}||\Xi_{i}||^{2} \Bigg | d_{1}, d_{2}, \cdots, d_{N}\Bigg]\leq E\Bigg[\sum^{N}_{i=1}||\Xi_{i}||^{2} \Bigg | d_{N}\Bigg]
\end{eqnarray}
where the equality holds if and only if $\delta_{j}\geq \delta_{N}$ for $\forall j\in \{1,2, \cdots, N-1\}$.
\end{theorem}
\textbf{Proof:} $[\Xi_{1}^{'},  \Xi_{2}^{'}, \cdots, \Xi_{N}^{'}]'\sim \mathcal{N}(\mathbf{0}, I_{Nn})$ since $\Xi_{j}\overset{\text{i.i.d}}{\sim}  \mathcal{N}(\mathbf{0}, I_{n})$ .
Now we switch to the spherical coordinates. 
\begin{align}\nonumber
&f_{\Xi_{1}, \Xi_{2}, \cdots, \Xi_{N}}(\xi_{1}, \xi_{2}, \cdots, \xi_{N})\mathrm{d}\xi_{1}\cdots \mathrm{d}\xi_{N}\\&=f(r, \Phi)r^{Nn-1}\mathrm{d}r\mathrm{d}\Phi=\frac{1}{U}f_{r}(r)r^{Nn-1}\mathrm{d}r\mathrm{d}\Phi
\end{align}
where  $r$ is the radius, $\Phi$ is the angle, and $U$ is the surface area of the $Nn$-dimensional unit ball. The second equality holds because the radius and the angle are independent for a Gaussian random vector.

Define $g(r)=r^2$ and $D=d_1\cap d_2 \cdots \cap d_N$. Then, by virtue of Lemma \ref{lem5}, one writes
\begin{eqnarray}
E\Bigg[\sum^{N}_{i=1}||\Xi_{i}||^{2} \Bigg | d_{1}, d_{2}, \cdots, d_{N}\Bigg]=
E[g(r)| D]=\frac{\int_{\Omega} \frac{1}{U}\int_0^{r_\Omega ^D} g(r)f_{r}(r)r^{Nn-1}  \mathrm{d}r\mathrm{d}\Phi}{\int_{\Omega} \frac{1}{U}\int_0^{r_\Omega ^D} f_{r}(r)r^{Nn-1}\mathrm{d}r\mathrm{d}\Phi}
\end{eqnarray}	
where $r_\Omega ^D$ is the distance from the origin to $\partial D$ at the angle $\Omega$. 

Let $P(x)=\int _0^x g(r)f(r)r^{Nn-1}\mathrm{d}r$, $Q(x)=\int _0^x f(r)r^{Nn-1}\mathrm{d}r$,
$R(x)=\frac{P(x)}{Q(x)}$. Then
\begin{align}\nonumber
\frac{ \mathrm{d}R(x)}{ \mathrm{d}x}&=\frac{g(x)f(x)x^{Nn-1}}{Q(x)}+\frac{-f(x)x^{Nn-1}\int _0^x g(r)f(r)r^{Nn-1}\mathrm{d}r}{Q^2(x)}\\\nonumber
&=\frac{f(x)x^{Nn-1}}{Q(x)}\Big(g(x)-\frac{\int _0^x g(r)f(r)r^{Nn-1}\mathrm{d}r}{Q(x)}\Big)\\
&\geq \frac{f(x)x^{Nn-1}}{Q(x)}\Big(g(x)-\frac{g(x)\int _0^x f(r)r^{Nn-1}\mathrm{d}r}{Q(x)}\Big)=0.
\end{align}	
Consequently, $R(x)$ is monotonically nondecreasing function of $x$.

Since $r_\Omega ^D\leq \sqrt{\delta_{N}}$, we have $R(r_\Omega ^D)\leq \frac{P(\sqrt{\delta_{N}})}{Q(\sqrt{\delta_{N}})}$ for $\forall \Omega$.
\begin{align}\nonumber
E[g(r)|D]=\frac{\int_\Omega P(r_\Omega^D)\mathrm{d}\Phi}{\int_\Omega Q(r_\Omega^D)\mathrm{d}\Phi}=\frac{\int_\Omega R(r_\Omega^D)Q(r_\Omega^D)\mathrm{d}\Phi}{\int_\Omega Q(r_\Omega^D)\mathrm{d}\Phi}\\
\leq \frac{\int_\Omega \frac{P(\sqrt{\delta_{N}})}{Q(\sqrt{\delta_{N}})}Q(r_\Omega^D)\mathrm{d}\Phi}{\int_\Omega Q(r_\Omega^D)\mathrm{d}\Phi}=\frac{P(\sqrt{\delta_{N}})}{Q(\sqrt{\delta_{N}})}=E[g(r)|d_n].
\end{align}	
Obviously, the equal sign holds if and only if $D=d_n$, i.e., $\delta_{j}\geq \delta_{N}$ for $\forall j\in \{1,2, \cdots, N-1\}$.\ \ \ \ \ \ \ \ \ \ \ \ \ \ \ \ \ \ \ \ \ \ \ \ \ \ \ \  \ \ \ \ \ \ \ \ \ \ \ \ \  \ \ \ \ \ \ \ \ \ \ \ \ \ \ \ \  \ \ \ \ \ \ \ \ \ \ \ \ \ \  \ \ \ \ \ \ \ \ \ \ \ \ \  \ \ \ \ \ \ \ \ \ \ \ $\Box$
 
Note that the random vector $[\Xi_{1}',  \Xi_{2}', \cdots, \Xi_{N}']'$ in the context of this paper refers to the cumulative innovation $\epsilon_{t+1,k}$. Theorem \ref{th2} reveals that the trace of the covariance of the cumulative innovation, conditioned on the joint set of the entire non-transmission regions, does not exceed the one conditioned only on the last non-transmission region. 

This theorem furnishes a consequential result in both conceptual and applied terms. Conceptually, the lack of previous thresholding information, serves to magnify the covariance of the cumulative innovation. Conversely, knowledge of the whole chain of thresholding values in the entire non-transmission regions suppresses the covariance, i.e. uncertainty, of the cumulative innovation. This result evokes the distinct intuition that more knowledge translates into higher confidence in estimation, thereby reducing uncertainties. In the practical sense, the computation of the covariance of the cumulative innovation, given the irregularity of the boundary in the joint set of non-transmission regions, poses a formidable task and one is tempted to consider a workable upper bound. In fact, this inequality will soon be summoned to deliver an upper bound of (the trace of) the MSE matrices at the remote center under our proposed scheme. The following theorem reads:

\begin{theorem}
\label{th3}
If $\gamma_{k}=0$,
\begin{eqnarray}\label{eq4.7}
&&\text{tr}(P_{k})\leq \text{tr}(P_{k}^{L})+\tau n\beta(\tau,\delta_{k})\lambda_\text{max}(A^{P}_{k}),
\end{eqnarray}
where $\beta(\tau,\delta_{k})=\frac{\text{Pr}(\chi^2_{\tau n+2}\leq \delta_{k})}{\text{Pr}(\chi^2_{\tau n}\leq \delta_{k})}$, $A^{P}_{k}=\text{diag}(A^{i}(P^{L}_{k-i|k-i-1}-P^{L}_{k-i})(A^{i})^{
\prime})$, $i\in \{\tau-1, \cdots, 0\}$.
\end{theorem}
\textbf{Proof:} 
If $\gamma_{k}=0$,
\begin{align}\nonumber
\text{tr}(P_{k})&=\text{tr}(P_{k}^{L})+\text{tr}\Bigg(\sum_{i=0}^{\tau-1}\eta_{i}(\delta_{t+1}^{k})A^{i}(P^{L}_{k-i|k-i-1}-P^{L}_{k-i})(A^{i})^{
	\prime}\Bigg)\\\nonumber
&= \text{tr}(P_{k}^{L})+\text{tr}(\eta(\delta_{t+1}^{k})A^P_{k})\\\nonumber
&\leq \text{tr}(P_{k}^{L})+\text{tr}(\eta(\delta_{t+1}^{k}))\lambda_\text{max}(A^{P}_{k})\\\nonumber
&=\text{tr}(P_{k}^{L})+\mathbb{E}\Bigg [\sum^{k}_{i=t+1}||\epsilon_{i}||^2\Bigg | \Omega_{t+1}, \cdots, \Omega_{k-1}, \Omega_{k}\Bigg ]\lambda_\text{max}(A^{P}_{k})\\\nonumber
&\leq \text{tr}(P_{k}^{L})+\mathbb{E}\Bigg [\sum^{k}_{i=t+1}||\epsilon_{i}||^2\Bigg | \Omega_{k}\Bigg ]\lambda_\text{max}(A^{P}_{k})\\
&=\text{tr}(P_{k}^{L})+\tau n\beta(\tau,\delta_{k})\lambda_\text{max}(A^{P}_{k})
\end{align}
where the second inequality holds due to Theorem \ref{th2}, and the last equality follows from Lemma 3 in \cite{bian2018remote}. \ \ \ \ \ \ \ \ \ \ \ \ \ \ \ \ \ \ \ \ \ \ \ \ \ \ \  \ \ \ \ \ \ \ \ \ \ \ \ \ \ \ \ \ \ \ \ \ \ \ \ \ \ \ \ \ \ \  \ \ \ \ \ \ \ \ \ \ \ \ \ \ \ \ \ \ \ \  \ \ \ \ \ \ \ \ \ \ \ \ \  \ \ \ \ \ \ \ \ \ \ \ $\Box$

Theorem \ref{th3} is the master result for the first part of this paper. 

One can read off the two error terms from it. The error at the remote center does not deviate significantly from the irreducible error stemming from the local estimate: the additional error introduced by the cumulative innovation-driven communication scheme is conveniently manageable due to its explicitly simple expression, implying that this upper bound is tight. This tight upper bound is readily deployable and may pave the way for a plurality of theoretical and practical studies that build on the platform of the cumulative innovation-driven communication scheme. For instance, this upper bound may enable potential constructs for a suboptimal estimation whenever the difficulty arises of calculating the exact MSE matrix under a general thresholding policy. To be more specific, let us remind ourselves that in our framework, Problem 1 frames $\text{tr}(P_{k})$ as a measure of estimation uncertainties. To accommodate for the challenges from numerical integrations, thanks to Theorem \ref{th3}, we may choose to relax this criterion in Problem 1 by considering the suboptimal problem with $\text{tr}(P_{k})$ replaced by its upper bound.

\vspace{1\baselineskip}

We have so far attempted to deliver results without specifying the thresholding policy under the framework of our proposed scheme. The formulation of the optimal problem are often forced to be relaxed to a suboptimal problem due to computational limits. In what follows, we offer a different strategy that respects the optimality, at the cost of sacrificing some, possibly minor, generalities, though. The second part of this paper is dedicated to this alternative approach.

\section{Reformulation of the optimization problem and implementation}
In this section, we show that Problem 1 with our communication scheme can be reformulated as an MDP problem under reasonable assumptions. Then an algorithmic approach is given based on the MDP model. Finally, we extend our results to the optimal average cost problem.

\begin{assumption}\label{asm3} 
For each time $k$, $\delta_{t+1}\geq \delta_{t+2}\cdots \geq \delta_{k}$.
\end{assumption}
\begin{assumption}\label{asm4} 
$P^{L}_{k|k-1}=\hat{P}$ and $P_{k}^{L}=\bar{P}$, for $\forall\ k\in \mathbb{Z}$, where
$\hat{P}$ is the unique positive definite solution of the algebraic Riccati equation
\begin{eqnarray} \label{eq5.1} 
&&\hat{P}=A(\hat{P}-\hat{P}C'(C\hat{P}C'+R)^{-1}C\hat{P})A'+Q,
\end{eqnarray}  
and $\bar{P}$ is given by 
\begin{eqnarray} \label{eq5.2} 
&&\bar{P}=\hat{P}-\hat{P}C'(C\hat{P}C'+R)^{-1}C\hat{P}.
\end{eqnarray} 
\end{assumption}

\begin{remark}\label{remark1} 
Assumption \ref{asm3} comes into play if an intuition applies: the more data packet drops accumulate consecutively, the lower the threshold should be to facilitate sending of a data packet at the next time instant. 
\end{remark}
\begin{remark}\label{remark2} 
Since the pair (A,C) is observable, the prior state covariance $P_{k|k-1}^{L}$ and the posterior covariance $P_{k}^{L}$ converge to the steady-state values $\hat{P}$ and $\bar{P}$, respectively, exponentially fast. Meanwhile, we consider an infinite horizon, and we omit the transient estimation process at the sensor side. As a result, we use Assumption \ref{asm4} throughout the optimization problem.
\end{remark}

Under Assumptions \ref{asm3}, $\epsilon_{t+1,k}\in \Omega_{k} \Rightarrow \epsilon_{t+1,k-1}\in \Omega_{k-1}, \cdots, \epsilon_{t+1,t+1}\in \Omega_{t+1}$, i.e., $\epsilon_{t+1,i}  \in \Omega_{i}$ is sufficient but not necessary for $\epsilon_{t+1,j}\in \Omega_{j} (i>j\geq t+1)$ since $||\epsilon_{t+1,i}||^{2}\geq ||\epsilon_{t+1,j}||^{2}$. Thus, if $\gamma_{k}=0$, then $\{\mathcal{I}\}_{0}^{k}=\{\{\mathcal{I}\}_{0}^{t}, \Omega_{t+1}, \dots, \Omega_{k-1}, \Omega_{k}\}=\{\{\mathcal{I}\}_{0}^{t}, \Omega_{k}\}$ since $\epsilon_{t+1,k}\in \Omega_{k}$ implies $\epsilon_{t+1,k-1}\in \Omega_{k-1}, \cdots, \epsilon_{t+1,t+1}\in \Omega_{t+1}$.

As a consequence, under Assumptions \ref{asm3} and \ref{asm4}, according to Lemma \ref{lem2} and (\ref{eq3.4.2}), $P_{k}$ in (\ref{eq4.4.2}) can be rewritten as
\begin{eqnarray}\label{eq5.3}
&&P_{k}=\bar{P}+\beta(\tau_{k}^{+},
\delta_{k})\sum_{i=0}^{\tau_{k}^{+}-1}A^{i}(\hat{P}-\bar{P})(A^{i})^{\prime}
\end{eqnarray}
where $\beta(0, \delta_{k})=0$ and $\sum_{i=0}^{-1}(\cdot)=0$.

\subsection{Stochastic DP model $(S, U, W)$}
We now reformulate Problem 1. We consider a discrete-time dynamical system. 
For all $k\in \{1,2, \cdots\}$, the system state is defined as $s_{k}=(\tau_{k-1}^{+}, \delta_{k-1})\in S$, where $S= \mathbb{Z}\otimes \mathbb{R}^{+}$ is the state space; the system control is defined as $u_{k}=\mu_{k}(s_{k})=\mu_{k}(\tau_{k-1}^{+}, \delta_{k-1})\in U$, where $U= \mathbb{R}^{+}$ is the control space; the disturbance is defined as $\gamma_{k}=\mathbbm{1}(||\epsilon_{k-\tau_{k-1}^{+},k}||^2> u_{k})\in W$, where $W=\{0, 1\}$ is the disturbance space. Under Assumption \ref{asm3}, $\gamma_{k}$ are characterized by probabilities $\text{Pr}(\cdot|s_{k}, u_{k})$ defined on $W$, where
 \begin{align}\label{eq5.2.1}
\text{Pr}(\gamma_{k}=0|(\tau_{k-1}^{+}=\tau^{+}, \delta_{k-1}=\delta), u_{k}=u)&=\left\{
\begin{array}{l}
		\text{Pr}(\chi^2_{n}\leq u) \ \ \ \ \ \ \ \ \ \ \ \text{if}\ \tau^{+}=0\\
                \frac{\text{Pr}(\chi^2_{(\tau^{+}+1)n}\leq u)}{\text{Pr}(\chi^2_{\tau^{+} n}\leq \delta)} \ \ \ \ \ \ \ \ \ \text{if}\ \tau^{+}> 0
                \end{array}
\right.\\\label{eq5.2.2}
\text{Pr}(\gamma_{k}=1|(\tau_{k-1}^{+}=\tau^{+}, \delta_{k-1}=\delta), u_{k}=u)
&=1-\text{Pr}(\gamma_{k}=0|(\tau_{k-1}^{+}=\tau^{+}, \delta_{k-1}=\delta), u_{k}=u).
\end{align}

Obviously, the state $s_{k}$ evolves according to the following stationary discrete-time dynamical system
\begin{eqnarray}\label{eq5.2.3}
&&s_{k+1}=f(s_{k}, u_{k}, \gamma_{k}), \ \ \ \ \ \ \ \ k=1, 2, \cdots
\end{eqnarray}
where $f: S\times U\times W \mapsto S$ and is given by
\begin{eqnarray}\label{eq5.2.4}
&&\left\{
\begin{array}{l}
\tau_{k}^{+}=(\tau_{k-1}^{+}+1)(1-\gamma_{k})\\
\delta_{k}=u_{k}.
\end{array}
\right.
\end{eqnarray}

We further define the admissible policy at time $k$ as $\mu_{k}=\{\mu \in S \mapsto U| \mu(\tau^{+}, \delta)\in U,\ \mu(\tau^{+}, \delta)\leq \delta\ \text{for}\ \tau^{+}> 0\}$ and the cost per stage $g: S\times U\times W \mapsto \mathbb{R}$ as 
\begin{eqnarray}\label{eq5.2.5}
g(s_k, u_k, \gamma_{k})=\text{tr}(\bar{P})+\beta(\tau_{k-1}^{+}+1,u_k)\text{tr}\Bigg(\sum_{i=0}^{\tau_{k-1}^{+}}A^{i}(\hat{P}-\bar{P})(A^{i})^{'}\Bigg)(1-\gamma_{k})+\kappa \gamma_{k}.
\end{eqnarray}
According to Theorem 2 in \cite{bian2018remote}, the cost per stage $g$ is bounded.

Given an initial state $s_1=(\tau_{0}^{+}, \delta_{0})$, we wish to find an admissible policy $\pi=\{\mu_1, \cdots, \mu_k, \cdots\}$ that minimizes the cost function
\begin{eqnarray}\label{eq5.2.6}
 &&J_{\pi}(s_1)=\underset{N \to \infty}{\text{lim}}\underset{\gamma_{k} \atop k=1,2, \cdots}{E}\Bigg[\sum_{k=1}^{N}\alpha^{k-1}g(s_k, \mu_k(s_k), \gamma_{k})\Bigg]
\end{eqnarray}
subject to the system (\ref{eq5.2.3}), where $0\leq \alpha< 1$ is the discount factor.

Let $\Pi$ be the set of admissible policies $\pi$. Then the optimal cost function $J^{*}$ is defined by
\begin{eqnarray}\label{eq5.2.7}
&&J^{*}(s)=\underset{\pi\in \Pi}{\text{min}}J_{\pi}(s),\ \ \ \ \ \  \ \ \ \ \ \  s\in S.
\end{eqnarray}

Note that if assumption \ref{asm3} and \ref{asm4} are satisfied, the differences between Problem 1 under our communication scheme and the optimization problem (\ref{eq5.2.7}) arise from the construct that the policy is Markovian and deterministic (MD), in contrast to the general decision rule which is history dependent and randomized (HR). It will be explained in Remark \ref{remark3} that the optimal cost under MD construct is equal to the one that uses HR.

Let 
\begin{eqnarray}\label{eq5.2.8}
&&(TJ)(s)=\underset{u\in U(s)}{\text{min}}\underset{\gamma_{k} \atop k=1,2, \cdots}{E}[g(s,u,\gamma)+\alpha J(f(s,u,\gamma))],\ \ \ \ \ \  \ \ \ \  s\in S.
\end{eqnarray}
Then the optimal cost function $J^{*}$ in $(\ref{eq5.2.7})$ satisfies Bellman's equation \cite{bertsekas1995dynamic} 
\begin{eqnarray}\label{eq5.2.9}
&&J^{*}(s)=(TJ^{*})(s).
\end{eqnarray}
Furthermore, $J^{*}$ is the unique solution of $(\ref{eq5.2.9})$ with the bounded function $g$.

\subsection{Finite-state MDP model $(S, U, p, g)$ and algorithmic design}

We further approximate the state space and control space over finite lattices\footnote[1]{For computational feasibility, we  constrain the state space to be $S= \{0,1, \cdots, M\}\otimes (0, \delta_{max})$ and control space $U= (0, \delta_{max})$.}, which enables us to formulate the optimization problem (\ref{eq5.2.7}) as a finite-state MDP problem $(S, U, p, g)$\footnote[2]{The solution to this MDP problem approaches the one for (\ref{eq5.2.7}) at the limits of unboundedness and continuum, i.e., $M\to \infty$ and $\zeta\to 0$. Therefore the solution is asymptotically optimal.}, where 

(1) State space $S=\{(\tau^{+}, \delta)| (\tau^{+}, \delta)\in \{0,1, \cdots, M\}\otimes \{\zeta, 2\zeta, \cdots, N\zeta\}\}$, where $\zeta$ is the lattice spacing and $N=\lceil \delta_{max}/\zeta \rceil$ 

(2) Control space $U=\{\zeta, 2\zeta, \cdots, N\zeta\}\}$

(3) The transition probability is stationary. Define $s=(\tau_{k-1}^{+}, \delta_{k-1})$ and $s'=(\tau_{k}^{+}, \delta_{k})$, according to (\ref{eq5.2.1}), (\ref{eq5.2.2}) and (\ref{eq5.2.3}), 
\begin{align}\label{eq5.3.1}
p_{ss'}(u)&=\text{Pr}(\tau_{k}^{+}=\tau^{+'}, \delta_{k}=\delta'|\tau_{k-1}^{+}=\tau^{+}, \delta_{k-1}=\delta, u_{k}=u)\\\nonumber
&=\delta_{u \delta'}\left\{
\begin{array}{l}
		\text{Pr}(\chi^2_{n}\leq u) \ \ \ \ \ \ \ \ \ \ \ \ \ \ \ \ \ \ \text{if}\ \tau^{+}=0, \tau^{+'}=1\\
		1-\text{Pr}(\chi^2_{n}\leq u) \ \ \ \ \ \ \ \ \ \ \ \ \ \text{if}\ \tau^{+}=0, \tau^{+'}=0\\
		0 \ \ \ \ \ \ \ \ \ \ \ \ \ \ \ \ \  \ \ \ \ \ \ \ \ \ \ \ \ \ \text{if}\ \tau^{+}=0, \tau^{+'} \in \{2, \cdots,M\}\\
		\mathbbm{1}(u\leq \delta)\left\{
\begin{array}{l}
                \frac{\text{Pr}(\chi^2_{(\tau^{+}+1)n}\leq u)}{\text{Pr}(\chi^2_{\tau^{+} n}\leq \delta)} \ \ \ \ \ \ \ \ \ \ \ \ \ \ \ \text{if}\ \tau^{+} \neq 0, \tau^{+'}=\tau^{+}+1\\
                	1-\frac{\text{Pr}(\chi^2_{(\tau^{+}+1)n}\leq u)}{\text{Pr}(\chi^2_{\tau^{+} n}\leq \delta)} \ \ \ \ \ \ \ \ \ \ \text{if}\ \tau^{+} \neq 0, \tau^{+'}=0\\
	         0 \ \ \ \ \ \ \ \ \ \ \ \ \ \ \ \ \  \ \ \ \ \ \ \ \ \ \ \ \ \text{if}\ \tau^{+} \neq 0, \tau^{+'}\neq \tau^{+}+1, \tau'\neq 0
		\end{array}
                \right.
\end{array}
\right.
\end{align}

(4) The expected cost per stage 
\begin{eqnarray}\label{eq5.3.2}
&&g(s, u)=\underset{s'}{\Sigma}p_{ss'}(u)\hat{g}(s,u,s')
\end{eqnarray}
where the cost of using $u$ at state $s$ and moving to state $s'$
\begin{eqnarray}
\hat{g}(s,u,s')=\left\{
\begin{array}{l}
\text{tr}(\bar{P})+(\beta(\tau^{+'}, u)\text{tr}(\sum_{i=0}^{\tau^{+'}-1}A^{i}(\hat{P}-\bar{P})(A^{i})^{'})\ \ \ \ \ \text{if}\ \tau^{+'}>0\\
\text{tr}(\bar{P})+\kappa\ \ \ \ \ \ \ \ \ \ \ \ \ \ \ \ \ \  \ \ \ \ \ \ \ \ \ \ \ \ \ \ \ \ \ \ \ \ \ \ \ \ \ \ \ \ \ \ \text{if}\ \tau^{+'}=0.
\end{array}
\right.
\end{eqnarray}

The mapping $T$ of (\ref{eq5.2.8}) can be written as
\begin{eqnarray}\label{eq5.3.3}
&&(TJ)(s)=\underset{u\in U(s)}{\text{min}}[g(s,u)+\alpha \underset{s'}{\Sigma}p_{ss'}(u)J(s')].
\end{eqnarray}

\begin{theorem}
\label{th4}
For the infinite horizon discounted cost MDP problem $(S, U, p, g)$, there exists an optimal stationary policy $\mu$ such that
\begin{eqnarray}\label{eq5.3.4}
&&J_{\mu}(s)=J^{*}(s)\ \ \ \ \ \ \ \ \ \ \ \ \ \text{for}\ \forall s\in S
\end{eqnarray}
where $J^{*}(s)$ is defined in (\ref{eq5.2.7}).
\end{theorem}
\textbf{Proof:}  The infinite horizon discounted cost MDP problem $(S, U, p, g)$ has a discrete state space $S$ and a finite control space $U$ for each $s\in S$. According to Theorem 6.2.10 in \cite{puterman1994markov}, an optimal deterministic stationary policy always exists that attains the optimal cost under $\Pi$. Therefore $J_{\mu}(s)$ must attain $J^{*}(s)$.  \ \ \ \ \ \ \ \ \ \ \ \ \  \ \ \ \ \ \ \ \ \ \ \ \ \ \ \ \ \ \ \ \ \ \ \ \  \ \ \ \ \ \ \ \ \ \ \ \ \ \ \ \ \ \ \ \ \ \ \ \ \ \ \ \ \ \ \ \ \ $\Box$
\begin{remark}\label{remark3} 
Let $\Pi^{MD}$ denote the set of Markovian and deterministic policies and $\Pi^{HR}$ the set of history dependent and randomized policies. In our MDP framework, we consider $\Pi^{MD}$.
Theorem 6.2.10  of \cite{puterman1994markov} claims that $\mu$ obtains the optimality under most general decision rule $\Pi^{HR}$. Since $\mu\in \Pi^{MD}$ and $\Pi^{MD}\subset \Pi^{HR}$, we have $J_{\mu}(s)=\underset{\pi\in \Pi^{MD}}{\text{min}} J_{\pi}(s)=\underset{\pi\in \Pi^{HR}}{\text{min}} J_{\pi}(s)=J^{*}(s)$, that is, the optimal cost under our MD construct is equal to the one that uses HR. 
\end{remark}

This theorem enable us to content ourselves with the search for an optimal stationary policy. We rewrite our goal in (\ref{eq5.2.7}) as
\begin{eqnarray}\label{eq5.3.5}
&&J^{*}(s)=\underset{\mu}{\text{min}} J_{\mu}(s).
\end{eqnarray}

We now introduce a theorem that serves as a foundation for our iterative algorithmic design.
\begin{theorem}
\label{th5}
1) The optimal cost function $J^{*}$ satisfies
\begin{eqnarray}\label{eq5.3.6}
&&J^{*}(s)=\underset{N\to \infty}{\text{lim}} (T^{N}J)(s)\ \ \ \ \ \ \ \ \ \ \ \ \ \text{for}\ \forall s\in S.
\end{eqnarray}
2) A stationary policy $\mu$ is optimal if and only if $\mu(s)$ attains the minimum in Bellman's equation (\ref{eq5.2.9}) for each $s\in S$; that is, 
\begin{eqnarray}\label{eq5.3.7}
&&TJ^{*}=T_{\mu}J^{*}
\end{eqnarray}
where $(T_{\mu}J)(s)=g(s,\mu(s))+\alpha \underset{s'}{\Sigma}p_{ss'}(\mu(s))J(s')]$. 
\end{theorem}
\textbf{Proof:} Part 1 is a direct result of Proposition 2.1 in \cite{bertsekas1995dynamic} since $J:\ S\mapsto \mathbb{R}$ is bounded.
Part 2 follows from Proposition 2.3 in \cite{bertsekas1995dynamic}.\ \ \ \ \ \ \ \ \  \ \ \ \ \ \ \ \ \ \ \ \ \ \ \ \ \ \ \ \ \ \ \ \ \ \ \ \ \ \ \ \ \  \ \ \ \ \ \ \ \ \ \ \ \ \  \ \ \ \ \ \ \  \ \ \ \ \ \ \ \ \ \ \ \ \ \ \ \ \ \ \ \ $\Box$

Consequently, we can use value iteration to get the optimal cost function and policy numerically.

\begin{definition} (Definiton 5.2.1 in \cite{bertsekas2012dynamic})
State $s'$ is accessible from state $s$ if there exists a stationary policy $\mu$ and an integer $k$ such that
\begin{eqnarray}
&&\text{Pr}(s_k=s'|s_0=s, \mu)>0.
\end{eqnarray}
\end{definition}

Our MDP framework invites the following property of degeneracy.

\begin{proposition}
\label{pro1} 
Given any states $(0, \delta)$, $\forall \delta\in \{\zeta, 2\zeta, \cdots, N\zeta\}$ and an optimal stationary policy $\mu^{*}$, then:

1) The state $(0, \delta)$ degenerates w.r.t $\mu^{*}$, i.e., $\mu^{*}(0, \delta)=\mu^{*}(\tau^{+}=0)$, which is independent of the value of $\delta$.

2) If $(\tau^{+}, \delta_{1})$, $(\tau^{+}, \delta_{2})$ are both accessible from $(0, \delta)$ under $\mu^{*}$ for $\forall \tau^{+}\in \{1, 2, \cdots, M\}$, then $\delta_{1}=\delta_{2}$.
\end{proposition}
\textbf{Proof:} Part 1 follows from the fact that the optimality equations for $J(\tau^{+}=0, \delta)$ have identical form for all $\delta$'s and, perforce, degenerate all $(\tau^{+}=0, \delta)$ for all $\delta$'s.

Part 2 comes from the observation that since any current state $(\tau^{+}=0)$ or $(\tau^{+}>0, \delta_{\tau^{+}})$ can only transition to $(\tau^{+}=0)$ or $(\tau^{+}+1, \delta_{\tau^{+}+1})$, where $\delta_{\tau^{+}+1}$ is determined by $\mu^{*}$ and current state, with nonzero probability, any accessible state $(\tau^{+}, \delta_{\tau^{+}})$ can  trace back to $(\tau^{+}=0)$ via a single path. \ \ \ \ \ \ \ \ \ \ \ \ \  \ \ \ \ \ \ \ \ \ \ \ \ \ \ \ \ \ \ \ \ \ \ \ \ \ \ \ \ \ \ \ \ \  \ \ \ \ \ \ \ \ \ \ \ \ \  \ \ \ \ \ \  \ \ \  \ \ \ \ \ \ \ \ \ \ \ \ \  \ \ \ \ \ \ \ \ \ \ \ $\Box$

This property essentially tells that under our MDP settings, the optimal stationary policy $\mu^{*}$, given any initial states $(\tau^{+}=0, \delta)$, confines the entire $(M+1)\times N$-dimensional state space to a $\tau^{+}$-indexed subspace of $(M+1)$-dimension, necessitating the second state parameter $\delta$ to collapse to a fixed $\delta_{\tau^{+}}$ for the same value of $\tau^{+}$.

\subsection{Expected average cost over infinite horizon}
We now consider the optimization problem of the expected average cost over the infinite horizon 
\begin{eqnarray} \label{eq5.4.1}
&&\underset{\pi^\gamma}{\min} \underset{T \to \infty}{\lim}\frac{1}{T} E^{\pi^\gamma}\bigg[\sum^{T}_{k=1}(\text{tr}(P_{k})+\kappa\gamma_{k})\bigg].
\end{eqnarray}

Following the lines of subsections A and B in section IV, one could model problem (\ref{eq5.4.1})  also in an MDP framework. It is desired to find a policy $\pi=\{\mu_1, \cdots, \mu_k, \cdots\}$ to minimize the cost function
\begin{eqnarray}\label{eq5.4.2}
 &&H_{\pi}(s_1)=\underset{T \to \infty}{\lim}\frac{1}{T} E\Bigg[\sum_{k=1}^{N}g(s_k, \mu_k(s_k))\Bigg].
\end{eqnarray}

Let $\Pi$ be the set of admissible policies $\pi$. Then the optimal cost function $H^{*}$ is defined by
\begin{eqnarray}\label{eq5.4.3}
 &&H^{*}(S)=\underset{\pi\in \Pi}{\text{min}}\ H_{\pi}(S).
 \end{eqnarray}
 
It is noteworthy that for arbitrary two states $s=(\tau^{+}, \delta)$ and $s'=(\tau^{+'}, \delta')$ in the dynamical system (\ref{eq5.2.3}), the accessibility condition holds. This observation helps deliver a conclusion of both theoretical and practical implications (Proposition 2.6 in \cite{bertsekas1995dynamic}): the optimal average cost per stage has the same value for arbitrary initial states if the accessibility condition holds and further,
\begin{eqnarray}\label{eq5.4.4}
&&\lambda=\underset{\alpha \to 1}{\text{lim}}\ (1-\alpha)J^{*}_{\alpha}(s)\ \ \ \ \ \ \ \forall s\in S
\end{eqnarray}
where $\lambda$ is the optimal average cost per stage and $J^{*}_{\alpha}$ is $J^{*}$ in (\ref{eq5.3.4}). That is, $\lambda=H^{*}(s)$, which is independent of the initial state $s$.

From a practitioner's point of view, one could, at least in principle, obtain the optimal average cost per stage and the corresponding optimal policy by taking advantage of the converging sequence of $(1-\alpha)J^*_{\alpha}$. It is the pathway we follow in our quest for the optimal average cost.

\section{Illustrative example}
This section presents an example to illustrate the results in this paper.

Consider the following stochastic linear system
\begin{align}\label{eq6.1}
x_{k+1}&=\left [
\begin {array} {cccc}
1&1\\
0&1 \end {array}\right ]x_{k}+w_{k}\\
y_{k}&=\left [
\begin {array} {cccc}
1&1\\
0&1.3 \end {array}\right ]x_{k}+v_{k}
\end{align}
with $Q=5I_{2}$, $R=2I_{2}$, $P_{0}=0.3I_{2}$, 
$x_{0}=[1,1]'$.

\begin{figure}[!htb]
\centering
\includegraphics[width=3.5in]{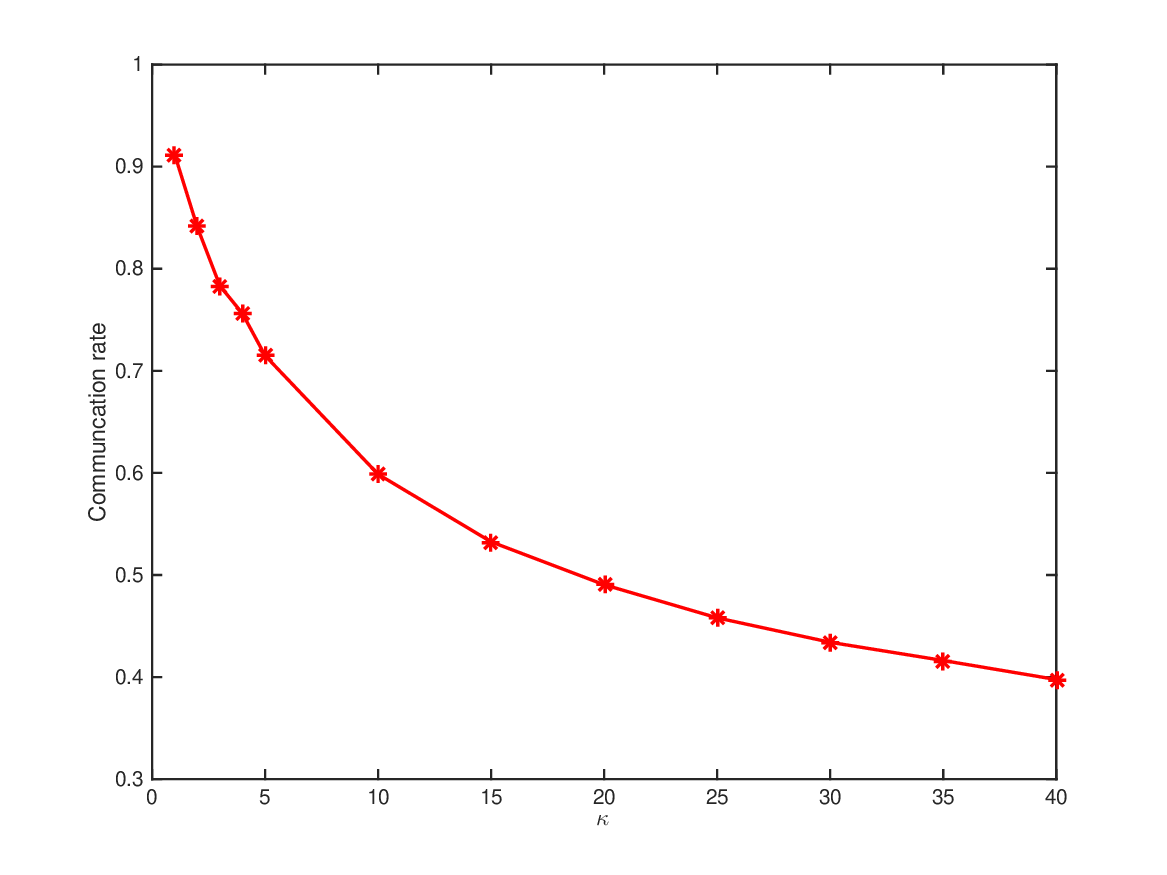}
\caption{Communication rate vs. $\kappa$}\label{fig1}
\end{figure}

Given the discount factor $\alpha=0.999$, the maximum elapsed time $M=6$ and the lattice spacing $\zeta=0.1$, by applying our MDP approach, the optimal discounted cost $J^{*}$, the optimal stationary policy $\mu^{*}$ and the approximated optimal average cost per stage $\hat{\lambda}=(1-\alpha)J^{*}$, versus different weighted unit transmission costs $\kappa=(5, \ 20, \ 35)$, are obtained and illustrated in Table \ref{table: 2}. $J^{*}$ and $H^{*}$ grow with an increase in the unit transmission cost $\kappa$. In addition, for the same value of $\tau^{+}$, the threshold rises with growing $\kappa$. The intuition behind the upshift is that a high expense of communication suppresses the rate of transmission. It harkens back to the goal of our policy that optimally balances communication consumption with estimation quality. Fig. \ref{fig1} depicts the optimal communication rates under different $\kappa$ ($1$ to $40$) by our MDP approach. The communication rates decrease with $\kappa$. It also illustrates the effect of our policy. 

\begin{table}[!hbp]
\caption{Optimization Results}\label{table: 2}
\centering
\begin{tabular}{|c|c|c|c|c|c|c|c|c|c|} 
\hline 
\multirow{2}{*}{$\kappa$}&\multirow{2}{*}{$J^{*}$}&\multicolumn{7}{|c|}{$U^{*}$}&\multirow{2}{*}{$H^{*}$}\\
\cline{3-9}
~&~&$\tau^{+}=0$&$\tau^{+}=1$&$\tau^{+}=2$&$\tau^{+}=3$&$\tau^{+}=4$&$\tau^{+}=5$&$\tau^{+}=6$&~\\
\hline
5&4152.8783& 0.9& 0.6 &0.4 &0.3 & 0.2 & 0.1 & 0.1&7.0679\\
\hline
20&12836.4439& 2.7&2&1.3&0.9& 0.6 & 0.4& 0.1&15.7515
\\
\hline
35&19607.0952
&4.1& 3 &2 & 1.3 & 0.9 & 0.7 &0.1&22.5221
\\
\hline
\end{tabular}
\end{table}

\begin{figure}[!htb]
\centering
\includegraphics[width=3.5in]{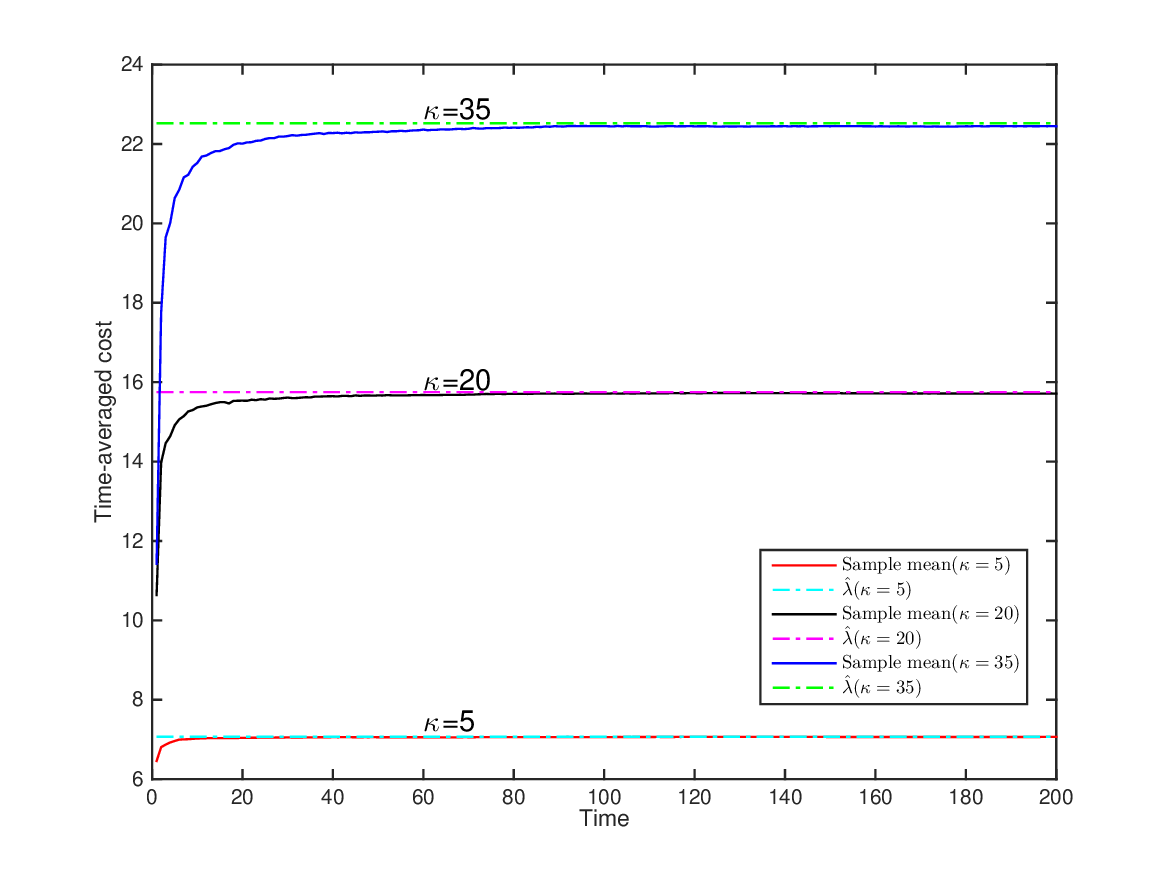}
\caption{Time averaged sample mean cost and $\hat{\lambda}$ for $\kappa=(5, \ 20, \ 35)$}\label{fig2}
\end{figure}

Define the time averaged sample mean cost as 
\begin{eqnarray} \label{eq6.2}
&&\frac{1}{T}\sum^{T}_{k=1}(\overline{\text{tr}(P_{k})+\kappa\gamma_{k}})
\end{eqnarray}
where $T$ is total time, $\overline{\text{tr}(P_{k})+\kappa\gamma_{k}}$ is the sample mean of cost per stage obtained from 500 Monte Carlo runs. Fig. \ref{fig2} compares (\ref{eq6.2}) with 
$\hat{\lambda}$  for $\kappa=(5, \ 20, \ 35)$ and shows that they all converge to $\hat{\lambda}$ fast with an increase in runtime. The validity of the approximated theoretical value is thereby manifested. 

\begin{figure}[!htb]
\centering
\includegraphics[width=3.5in]{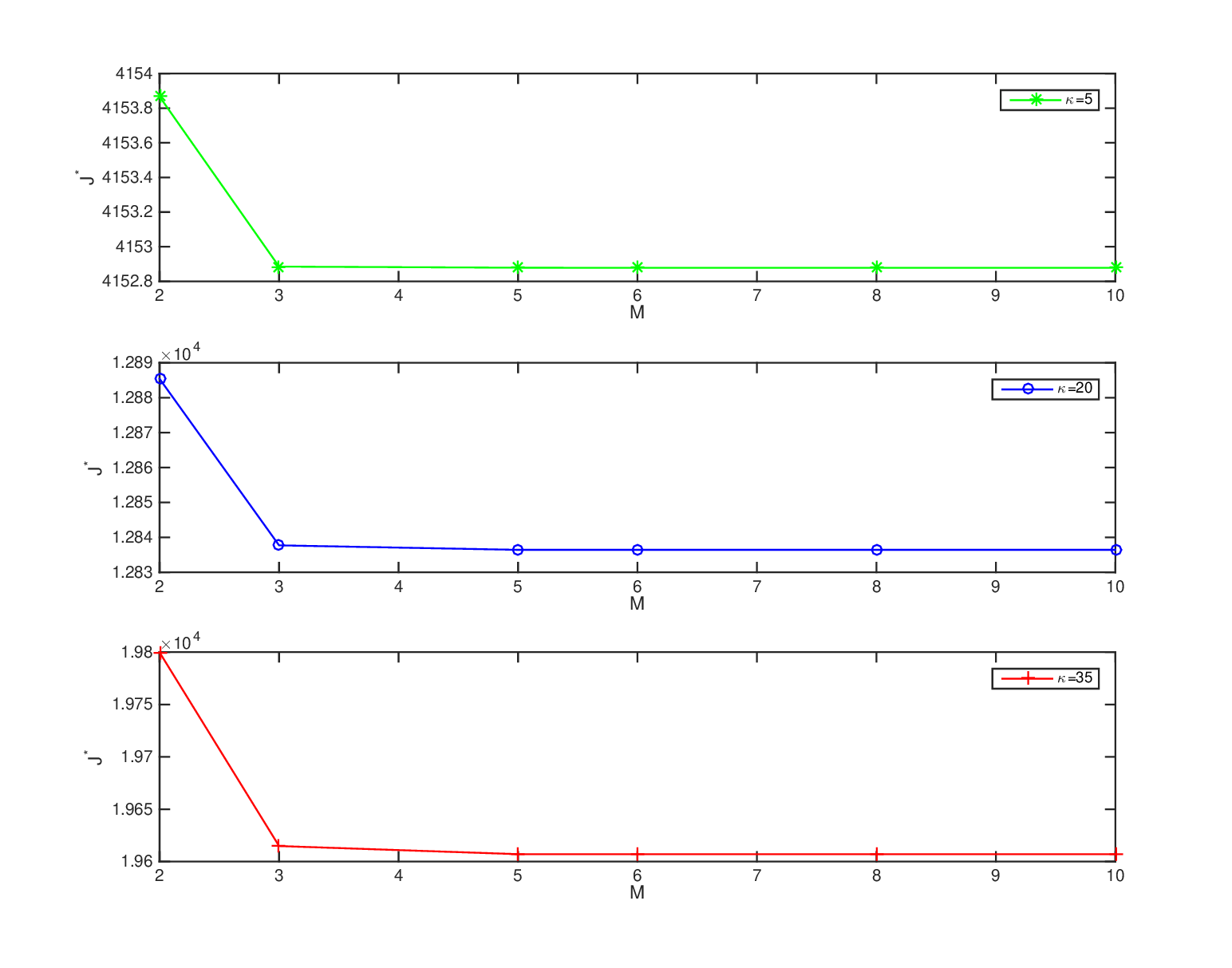}
\caption{$J^{*}$ for $\kappa=(5, \ 20, \ 35)$ vs. $M$}\label{fig3}
\end{figure}

\begin{figure}[!htb]
\centering
\includegraphics[width=3.5in]{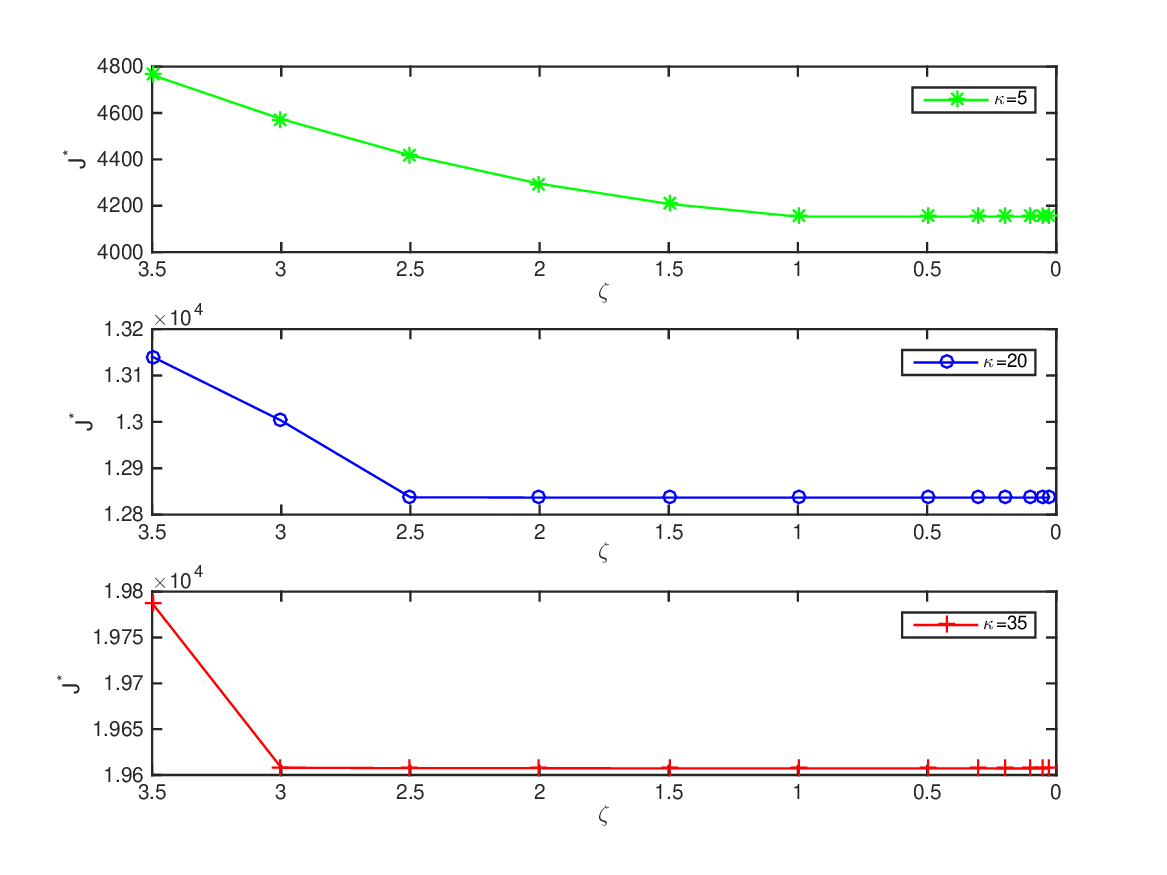}
\caption{$J^{*}$ for $\kappa=(5, \ 20, \ 35)$ vs. $\zeta$}\label{fig4}
\end{figure}

For $\kappa=(5, \ 20, \ 35)$, Fig. \ref{fig3} and Fig. \ref{fig4} depict $J^{*}$ as a function of $M$ and $\zeta$, respectively. Fig. \ref{fig3} shows that for smaller $M$, $J^{*}$ drops. But it converges even at moderately large $M$. Therefore, the value $6$ we used before already works with great precision. It is observed that the trend plateaus faster for a smaller $\kappa$, which corroborates the fact that our optimal policy for a smaller $k$ tends to suppress the probability of large $\tau^{+}$ and therefore the influence of $M$ on $J^{*}$ is marginal. This effect is also clear from Table \ref{table: 2}---at $\kappa=5$, $U^{*}$ for $\tau^{+}=5$ already drops to $0.1$ (the smallest possible threshold). Fig. \ref{fig4} shows the influence of lattice spacing---$J^{*}$ drops for smaller $\zeta$; yet it converges at least at $\zeta=0.5$. This observation enables us to use the value $0.1$ with high confidence. One should notice that the effect of varying $\zeta$ is more pronounced for a smaller $\kappa$. That is because smaller $\kappa$ corresponds to a lower threshold (shown in Table \ref{table: 2}) for the same $\tau^{+}$, thereby sensitizing the choice of $\zeta$.

\begin{figure*}[!htb]
\begin{center}
\subfigure[Discounted costs vs. $\kappa$]{
\includegraphics[width=3in]{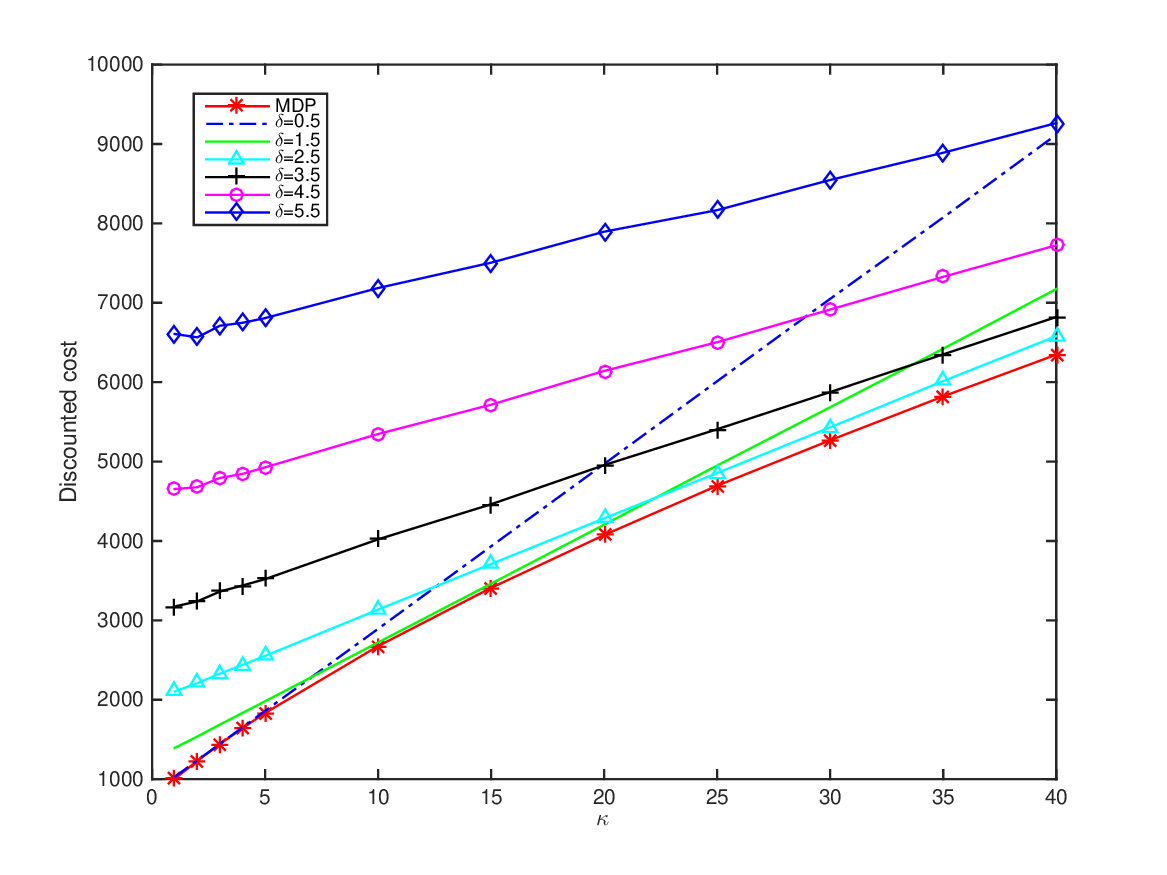}
}\hspace{-10mm}
\subfigure[Time averaged sample mean costs vs. $\kappa$]{
\includegraphics[width=3in]{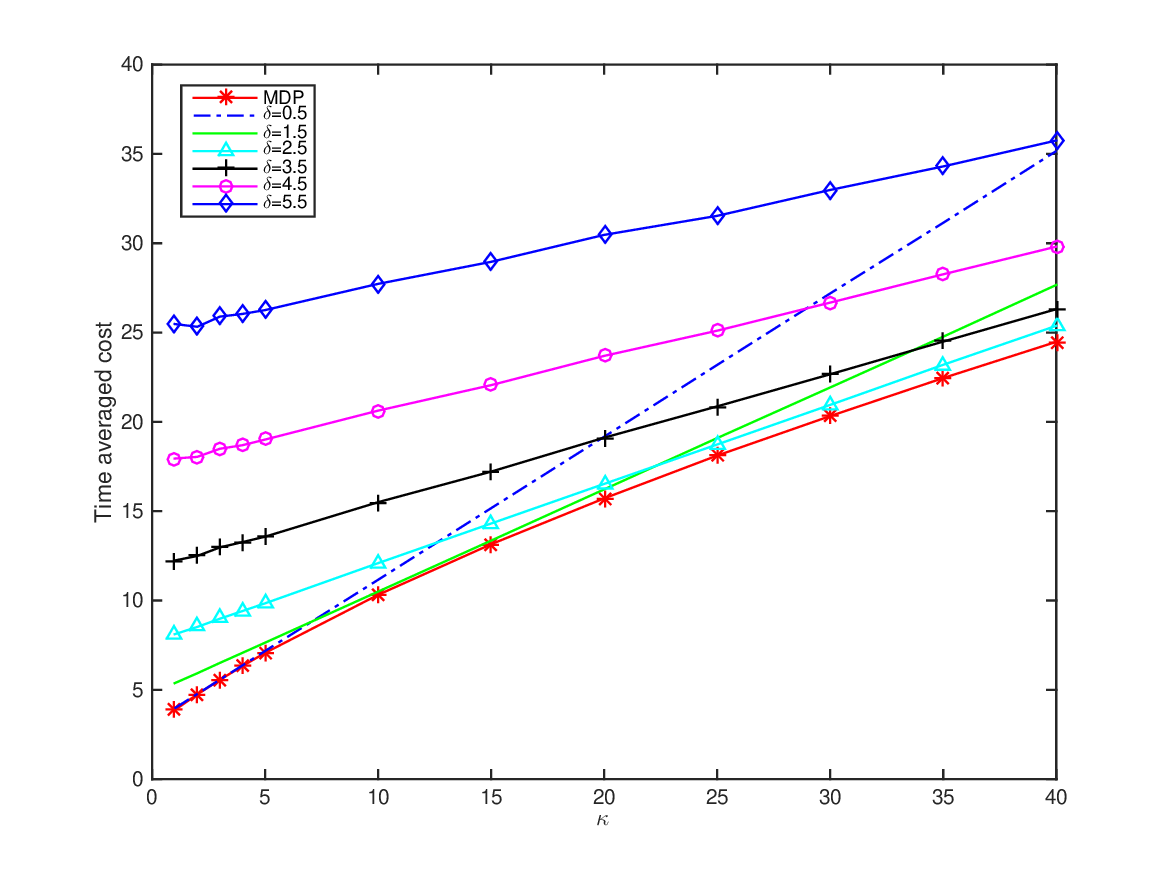}
}
\caption{Discounted costs and time averaged sample mean costs for MDP policy and six fixed thresholding policies with $\delta=0.5:1:5.5$ vs. $\kappa$}\label{fig5}
\end{center}
\end{figure*}

For our MDP approach and six other fixed thresholding policies with $\delta=0.5:1:5.5$, we compare their discounted costs and time averaged sample mean costs for different $\kappa$ values over 400 time steps and 500 Monte Carlo runs in Fig. \ref{fig5} (a) and (b), respectively. They both confirm that our optimal policy produces the minimal costs for different $\kappa$. Some fixed thresholding policies exhibit close behaviors at some small regions of $\kappa$, but it is our policy that consistently yields the best result throughout the entire spectrum of $\kappa$. That is because our MDP methodology builds on solid theoretical grounds that ensures the asymptotic optimality of the solution through a systematic procedure.

\section{Summary and conclusions}
This paper focuses on the optimality problem of an ETC scheme and a state estimator to quantitatively balance system performance and limited communication resources. We propose an ETC scheme based on the cumulative estimate innovation and time-varying thresholds and give its transmission probability. We derive the corresponding remote MMSE estimator. We present a workable tight upper bound for the MSE matrices. We define the optimization criterion based on the expected total discounted cost over the infinite horizon and show that our system is of an MDP nature, enabling an iterative algorithm to find the optimal cost and optimal policy. We also extend our results to solving the optimal average cost problem. We prove that our framework satisfies DOF reduction; specifically, the current optimal thresholding policy depends on no other state variables other than the tally of the elapsed time since the last transmission. Our simulation results illustrate the effectiveness of our approach.

\printbibliography

\end{document}